\documentstyle[12pt,epsfig]{article}
\newcommand{\beq}[1]{\begin{equation}\label{#1}}
\newcommand{\eeq}{\end{equation}}
\newcommand{\beqar}[1]{\begin{eqnarray}\label{#1}}
\newcommand{\eeqar}{\end{eqnarray}}
\newcommand{\lash}[1]{\not\! #1 \,}
\newcommand{\bra}[1]{\big< #1 \big|}
\newcommand{\ket}[1]{\big| #1 \big>}
\newcommand{\nn}{\nonumber}
\newcommand{\ep}{\varepsilon}

\newcommand{\La}{\Lambda}

\begin{document}
\vspace{2cm}
\begin{center}
{\LARGE \bf 
The Chiral-Odd Distribution Function $h_1$:\\[2mm]
IR-Renormalon Contribution\\[3mm]
and $\alpha_s$-Correction
}
\vspace{1cm}

M.~Meyer-Hermann$^1$, 
and A.~Sch\"afer$^2$
\vspace{1cm}

$^1$Institut f\"ur Theoretische Physik, J.~W.~Goethe
Universit\"at Frankfurt,
\\ 
Postfach~11~19~32, D-60054~Frankfurt am Main, Germany
\\ \vspace{2em}
$^2$Fakult\"at Physik, Universit\"at Regensburg, D-93040~Regensburg,
Germany
\end{center}

\vspace{2cm}
\noindent{\bf Abstract:}
The chiral-odd distribution function $h_1$ is discussed in the
framework of the renormalon approach. Using a bag-model calculation
for $h_1$, we calculate its intrinsic
uncertainty due to renormalon poles. The result
is given as a function of Bjorken-$x$ as well as
for the first moments separately.
We also extract the perturbative corrections to the first
moments of $h_1$.
\vspace*{\fill}
\eject
\newpage

\section{Introduction}

While in totally inclusive
deep inelastic scattering (DIS) the quark chirality is conserved
up to terms proportional to the quark masses, this is not the case in 
the Drell-Yan process. Here a quark-antiquark-pair is annihilated to
a virtual photon, so that in the cross-section
the quarks originating and ending in the
same nucleon may carry different chirality.
This gives raise to chiral odd distribution functions, which first
appeared in the discussion of the transverse polarized Drell-Yan process
\cite{Ral79}. This chiral odd distribution function is defined by
a twist-2 operator and is called the transversity distribution
$h_1$. 
Unlike the helicity asymmetry $g_1$,
$h_1$ has no partonic interpretation in the chiral basis.
Changing to
the transversity basis \cite{Gol76} $g_1$ looses its partonic
interpretation and $h_1$ gets one instead. It is interpreted as
the probability to find a
quark in a transversely polarized nucleon in an eigenstate of the
transverse Pauli-Lubanski vector with eigenvalue $+1/2$, minus 
the same with eigenvalue $-1/2$.

Some experiments are planned to measure $h_1$ in the near future,
especially at RHIC (BNL)
and possibly by the COMPASS experiment at CERN
--- for a general review of the possibilities for measuring
the transversity distribution see \cite{Jaf963} ---, 
so that there is
need for theoretical predictions for $h_1$.
The anomalous dimension which we will also consider
in this contribution was calculated in \cite{Art90}.
In the nonrelativistic quark models $g_1$ and $h_1$ are identical.
The positivity of parton probabilities implies the inequality
$\left| h_1(x)\right|\leq f_1(x)$ and the
Soffer inequality \cite{Sof95}
$2\left| h_1(x)\right|\leq f_1(x)+g_1(x)$.
A bag model calculation predict
$\left|h_1(x)\right|\geq\left|g_1(x)\right|$, which
may be correct in general \cite{Jaf91}.
For medium large Bjorken-$x$ there exists a QCD-sum rule calculation
\cite{Iof95}, predicting a much smaller value for $h_1$ than the
already mentioned bag model calculation.

We wish to contribute to this theoretical discussion by calculating
an intrinsic uncertainty of the perturbative series due to
IR-renormalon poles. This uncertainty is a systematic error for the
experimental measurement of $h_1$.
In addition we discuss the anomalous dimension in the one loop
approximaion and extract
the $\alpha_s$ correction to the first moments of
the transversity distribution.

%===========================================================

\section{Definitions}

\subsection{Forward-Scattering-Amplitude}
The definition of $h_1$ in terms of an operator matrix element reads
\cite{Ral79,Jaf91,Iof95}:
\beqar{def}
\frac{i}{2} \int \frac{dy_+}{2\pi}\,e^{iy_+ x}
\bra{p,s} \overline{\psi}(0) \sigma_{\mu\nu} \gamma_5 \psi(y_+ n)
\ket{p,s}
&=&
  h_1(x,q^2)\left(s_{\bot\mu}p_\nu - s_{\bot\nu}p_\mu\right) \nn\\
+\, h_L(x,q^2)m^2\left(p_\mu n_\nu - p_\nu n_\mu\right) (s\cdot n)
&+& h_3(x,q^2)m^2\left(s_{\bot\mu} n_\nu - s_{\bot\nu} n_\mu\right)
\eeqar
where $x=-q^2/(2p\cdot q)$ is the Bjorken variable,
$n^\nu$ is a light cone vector with $n^2=0$ of dimension
$(mass)^{-1}$, $p$ is the proton momentum and $s$ is the proton spin.
$p^2=m^2$, $s^2=-1$ and $p\cdot s=0$. $p\cdot n=1$ and the transverse
part of the spin vector is defined by the decomposition
$s_\mu = (s\cdot n)p_\mu + (s\cdot p)n_\mu + s_{\bot\mu}$.
The contraction of this expression with the light cone vector
$n^\nu$ gives:
\beq{defn}
\hspace*{-1mm}
\frac{i}{2} \int \frac{dy_+}{2\pi}\,e^{iy_+ x}
\bra{p,s} \overline{\psi}(0) \sigma_{\mu\nu} n^\nu \gamma_5 \psi(y_+ n)
\ket{p,s}
\;=\;
h_1(x,q^2)s_{\bot\mu} - h_L(x,q^2)m^2 n_\mu (s\cdot n)
\quad.
\eeq
It is possible to relate the transversity distribution $h_1$ to the
imaginary part of the forward scattering amplitude on the leading
twist-level \cite{Iof95}:
\beq{Tmu}
T_\mu
\;=\;
\frac{i}{2}
\int d^4y\,e^{iqy}
\bra{p,s} T\left(j_{\mu 5}(y) j_S(0) + j_S(y) j_{\mu 5}(0) \right)
\ket{p,s}
\quad.
\eeq
Here $j_{\mu 5}(y) = \overline{\psi}(y) \gamma_\mu \gamma_5 \psi(y)$
is a axial-vector current,
$j_S(y) = \overline{\psi}(y) \psi(y)$ is a
scalar current and $T$ denotes the time-ordered product.
Summation over flavor indices is assumed.
%It shall be mentioned that the relation holds approximately
%also for singlet currents up to $\alpha_s$ and mass corrections,
%as the down quark contributions are strongly suppressed with
%respect to the up quark contributions \cite{Iof95}.
Equivalently one may use a time-ordered product of a vector-current
$j_{\mu}(y) = \overline{\psi}(y) \gamma_\mu \psi(y)$
and a pseudoscalar current
$j_5(y) = \overline{\psi}(y) \gamma_5 \psi(y)$.

To prove the relation between this definition and the
operator-definition Eq.\,(\ref{def})
of the transversity distribution, one has 
to decompose the forward scattering amplitude into its
Lorentz-structures using the conservation of either the axial-vector
or the vector current. The conservation of the axial-vector
current is correct only for the flavor nonsinglet current, so that
the proof remains correct up to an order $\alpha_s$-correction only.
As the vector-current is conserved independently of the flavor
combination under consideration
we prefer the definition
\beq{Tmu2}
T_\mu
\;=\;
\frac{i}{2}
\int d^4y\,e^{iqy}
\bra{p,s} T\left(j_{\mu}(y) j_5(0) + j_5(y) j_{\mu}(0) \right)
\ket{p,s}
\quad.
\eeq
To find the realation to $h_1$ 
the current product has to be expanded collecting the terms
with one incoming quark, one outgoing quark and one quark propagator
$S(y)=i\lash{\partial}\Delta(y)$, where $\Delta$ is the Pauli-Jordan
function. The Pauli-Jordan function is expanded on the light cone and
only the leading term is taken into account. The imaginary part of
the s-channel-term of the forward scattering amplitude takes
the form:
\beqar{Tmulc}
{\rm Im}\,T_\mu
&=&
-\frac{1}{2\pi}
\int d^4y\,e^{iqy}
\bra{p,s} :\overline{\psi}(y) \sigma_{\mu\nu} y^\nu i\gamma_5
\psi(0):
\ket{p,s} \delta'(y^2)
\nn\\
&& +\frac{1}{2\pi}
\int d^4y\,e^{iqy}
\bra{p,s} :\overline{\psi}(0) \sigma_{\mu\nu} y^\nu i\gamma_5
\psi(y):
\ket{p,s} \delta'(y^2)
\quad.
\eeqar
Here $\sigma_{\mu\nu}=\frac{i}{2}\left[\gamma_\mu,\gamma_\nu\right]$ and
$\delta'(y^2)=\left(\partial / \partial(y^2)\right) \delta(y^2)$.
Substituting light cone variables $y_-=y_0-y_3$ and $2y_+=y_0+y_3$,
integrating by parts, using translation invariance,
and putting this expression on the light cone by choosing $y$ such
that $y^2=0$, one gets
\beqar{Tmulc2}
{\rm Im}\,T_\mu
&=&
-\frac{i}{2}
\int \frac{dy_+}{2}\,e^{iy_+ x}
\bra{p,s} :\overline{\psi}(0) \sigma_{\mu\nu} n^\nu \gamma_5 
\psi(y_+ n):
\ket{p,s}
%\nn\\
%&& -\frac{i}{2}
%\int \frac{dy_+}{2}\,e^{-iy_+ x}
%\bra{p,s} :\overline{\psi}(0) \sigma_{\mu\nu} n^\nu \gamma_5
%\psi(y_+ n):
%\ket{p,s}
\quad.
\eeqar
This result has the form of Eq.\,(\ref{defn}),
so that one gets a relation between
the s-channel-part of the forward scattering amplitude
and the transversity
distribution by collecting the terms
proportional to $s_{\bot\mu}$:
\beq{Teqh1}
h_1(x) s_{\bot\mu} 
\;=\;
\frac{1}{\pi}\,
\left. {\rm Im}\, T_\mu \right|_{s_{\bot\mu}}
\quad.
\eeq
%
%The second term in Eq.\,(\ref{Tmulc2})
%differs from the first one by an additional sign in
%the exponential. They correspond to the exchange graphs on one hand
%and to the antiquark transversity contribution on the other hand. So
%the complete forward scattering amplitude --- again reduced to the terms
%proportional to $s_{\bot \mu}$ --- gives an antisymmetric
%combination of quark and
%antiquark transversity distributions:
%
%\beq{Teqh+h1}
%\left(h_1(x)+h_1(-x)\right) s_{\bot\mu} 
%\;=\;
%\left.\frac{1}{\pi}\, {\rm Im}\, T_\mu \right|_{s_{\bot\mu}}
%\eeq
%

%================================================================

\subsection{Light Cone Expansion and Moments}

The light cone expansion of the forward scattering amplitude may be
writen as:
\beq{Tmuom1}
\left. T_\mu \right|_{s_{\bot\mu}}
\;=\;
2 s_{\bot\mu}
\sum_{m=0}^\infty
C_m\left(\frac{Q^2}{\mu^2},\alpha_s\right) A_m(\mu^2)
\,\omega^{m+1} + {\rm higher\,twist}
\eeq
where $\omega=1/x$. $A_m$ are the reduced matrix elements of the
local twist-2 operators relevant for $h_1$:
\beq{rme}
\bra{p,s} \overline{\psi} \sigma^{\lambda\{\rho} i\gamma_5
iD^{\mu_1} \cdots iD^{\mu_m\}} \psi \ket{p,s}
\;=\;
2 A_m (s^\lambda  p^{\{\rho} - p^\lambda s^{\{\rho})
p^{\mu_1} \cdots p^{\mu_m\}} \psi - {\rm traces}
\quad.
\eeq
The brackets denote total symmetrization of all included
indices. The symmetrization and the subtraction of the traces 
are necessary to extract the leading twist part of the matrix element.
The moments of $h_1$ have to be defined as $M_m=C_m A_m$ to get the
general expression:
\beq{mom1}
2s_{\bot\mu} M_n 
\;=\;
\frac{1}{\pi} \int_0^1 dx\,x^n
\left.\left( Im\,T_\mu(x) + (-)^n Im\,T_\mu(-x)\right)
\right|_{s_{\bot\mu}}
\quad.
\eeq
Using Eq.\,(\ref{Teqh1}) the moments become:
\beq{mom2}
M_n
\;=\;
\frac{1}{2}
\int_0^1 dx\,x^n
\left( h_1(x) + (-)^n h_1(-x)\right)
\eeq
for $n=0,1,\ldots$
This result coincides with the expression found by Jaffe and Ji
\cite{Jaf91}.
%If one had used Eq.\,(\ref{Teqh+h1}) instead, that is to have
%inserted the full forward scattering amplitude including the exchange
%graphs, the result for the moments would have been the same
%except an additional factor 2 and a 
%restriction on even moments only. 
%Then it would be impossible to get
%information about odd moments using $T_\mu(x)$ instead of
%$T_\mu^s(x)$. Therefore in the following
%we will calculate the forward
%scattering amplitude without exchange graphs.

The operators in Eq.\,(\ref{rme}) have a well defined behaviour under
charge conjugation: they are $C$-odd for even $n$ and $C$-even for odd
$n$. This change in sign corresponds to the relative sign of $h_1(x)$ and
$h_1(-x)$ in the moments Eq.\,(\ref{mom2}).
To obtain the correct relation between the antiquark and the
quark transversity distribution one may
look at the first moment:
\beq{mom3}
M_0
\;=\;
\frac{1}{2}
\int_0^1 dx\,
\left( h_1(x) + h_1(-x)\right)
\quad.
\eeq
As this expression is odd under charge conjugation, the
antiquark transversity distribution should be
$\overline{h}_1(x)=-h_1(-x)$. It follows that the contributions of
sea quarks cancel for even moments in general and only the valence
quarks contribute, while the sea quark contributions add
for odd moments.

%==================================================================

\section{Renomalon ambiguity of $h_1$}

We are calculating the IR-renormalon contribution \cite{Mue85}
to the structure
function defined by Eq.\,(\ref{Tmu}) and (\ref{Teqh1}).
As its twist-2 part
is related to the transversity distribution $h_1$, the IR-renormalon
contribution may be interpreted as intrinsic ambiguity of the
perturbative expansion to $h_1$.
Perturbative series in QCD are asymptotic ones, implying that the
radiative corrections of higher and higher orders get 
smaller only up to a finite order $m_0$ and diverge 
for larger orders in $\alpha_s$.
%As a consequence the series has to be understood as an asymptotic
%series, which has to be truncated at the order $m_0$ 
%corresponding to the smallest term.
The uncertainty of the series is of the order
of the smallest contribution
%which is essentially related to large
%scale effects, becoming important for lower values of $Q^2$. Indeed,
and has the form of a power correction \cite{Mue93}, 
so that the QCD perturbative series on the lowest twist level
may be written as ($a_s=\alpha_s/(4\pi)$):
\beq{truncate}
C_n\left(Q^2=\mu^2,a_s\right)
\;=\;
\sum_{k=0}^{m_0} B_n^{(k)} a_s^k
+
C_n^{(1)} \frac{\Lambda_C^2 e^{-C}}{Q^2}
+
C_n^{(2)} \frac{\Lambda_C^4 e^{-2C}}{Q^4}
+
{\cal O}\left(\frac{1}{Q^6}\right)
\quad.
\eeq
Note that the $1/Q$ term is not present in the above expression. It
was shown for Drell-Yan on the one gluon exchange level
that these corrections are cancelled by
higher order perturbative contributions \cite{Ben95} and that the
$1/Q^2$-term is the leading power correction.
The uncertainty may be determined by the exact calculation of the
perturbative corrections up to the order $m_0(Q^2)$, which is a very
demanding procedure.

Instead we calculate
the forward scattering amplitude Eq.\,(\ref{Tmu}) on the
one gluon exchange level, using a Borel transformed effective gluon
propagator \cite{Ben93}:
\beq{gluon}
{\cal B}_{1/a_s}[a_s D^{ab}_{\mu\nu}(k)](u)
\;=\;
\delta^{ab}\frac{ g_{\mu\nu} - \frac{k_\mu k_\nu}{k^2}}{k^2}
\left(\frac{\mu^2 e^{-C}}{-k^2}\right)^{\beta_0 u}
\quad.
\eeq
$C$ corrects for
the renormalization scheme dependence ($C=-\frac{5}{3}$ for
$\overline{MS}$-scheme), $\mu$ is the renormalization scale,
and $u$ is the Borel parameter.
The effective gluon
propagator is constructed by replacing the coupling $a_s$ by the
running coupling constant, that is by a resummation of all quark- and
gluon-loop insertions in one gluon-propagator.
In the first order of $a_s$ this
propagator leads to exact results. Looking at higher order
corrections the restriction on one exchanged
effective gluon corresponds to the large $N_f$-limit
\cite{Vas81}, where $N_f$ denotes the number of quark flavors. 
The next-to-leading $N_f$-terms are
approximated by naive-nonabelianization (NNA) 
\cite{Bro95}. 
This corresponds to the replacement of the one loop QED-beta-function
by the QCD-beta-function $\beta_0 = 11 - \frac{2}{3} N_f$ 
or equivalently to $N_f \rightarrow N_f - \frac{33}{2}$.
The quality of this
approximation has been checked  \cite{Ste96,Mey96}
for the unpolarized
structure functions $F_2$ and $F_L$ and for the polarized structure
function $g_1$ by comparing the
NNA perturbative coefficients with the known
exact ones. This comparison
gave very reasonable results for $F_L$ and $g_1$,
while the results for $F_2$ are less convincing.

Formally, asymptotic freedom is destroyed in the large 
$N_f$-limit. One should recognize that the large $N_f$-limit is used
to select graphs and has to be understood as a definition of an
approximation procedure. At the end $N_f$ will be set to 4, so that
$\beta_0$ stays in an asymptotic free region. This procedure is
technically analogous to the use of the large $N_c$-limit \cite{tHo74},
even if the physical content is different.

In the Borel plane the ambiguity of the truncated perturbative series in
Eq.\,(\ref{truncate}) is reflected in IR-renormalon poles, which
hinder an unambigous inverse Borel transformation. 
This ambiguity of the perturbative series may be interpreted as
twist-4 contribution of the structure function defined by
Eq.\,(\ref{Tmu}) and (\ref{Teqh1}). Such estimations of higher twist
corrections gave very reasonable results
\cite{Das96,Dok96,Ste96,Mey96}.
However the transversity distribution is defined by pure twist-2
operators and the relation of Eq.\,(\ref{Tmu})
to the operator definition of the transversity
distribution Eq.\,(\ref{def})
was shown for the twist-2 part of the structure function
only. This means that the IR-renormalon ambiguity of the
structure function has to be interpreted as intrinsic uncertainty
of the perturbative expansion of the transversity
distribution $h_1$.
%The contribution from IR-renormalons cancel
%with the contributions of UV-renormalons on a genuine twist-4 level
%calculation, as was shown in general \cite{Big94} and verified for
%special cases \cite{Ste96,Mey96}. This gives a
%connection between IR-renormalons and the
%higher twist corrections. Although the relative magnitude 
%of this part and the
%full higher twist corrections is unknown a priori,
%the estimations made until today gave very reasonable results
%\cite{Das96,Dok96,Ste96,Mey96}.

%Nevertheless it is still possible to
%give an estimate of the higher twist contributions
%in the framework of IR-renormalon method.
%It is exactly the philosophy of this method, to use exclusively
%the properties of the twist-2 perturbative series to make predictions
%for the twist-4-part.
So let us calculate the Borel-transformed s-channel
forward scattering amplitude $T_\mu$ in
Eq.\,(\ref{Tmu}) 
on the one gluon exchange level using the effective gluon
propagator Eq.\,(\ref{gluon}).
The result is a series in $\omega=1/x$
which has to be compared with Eq.\,(\ref{Tmuom1})
\beq{Tborel}
\left. {\cal B}_{\frac{1}{a_s}}
\left( T_\mu \right)\right|_{s_{\bot\mu}}
\;=\;
s_{\bot\mu}
\sum_{m=0}^\infty
{\cal B}_{\frac{1}{a_s}} 
\left(C_m\left(\frac{Q^2}{\mu^2},a_s\right)\right)  A_m(\mu^2)
\,\omega^{m+1} + {\rm higher\,twist}
\eeq
where the reduced matrix element was determined at the tree level to
be $A_n=1$ and a factor $2$ is missing because the exchange graphs
are not included in this expression due to the restriction on the
s-channel contribution. We obtain for the Borel transformed Wilson
coefficient:
\beqar{wilson}
&&\hspace*{-1cm}{\cal B}_{\frac{1}{a_s}}
\left[C_n\left(\frac{Q^2}{\mu^2},a_s\right)\right](s)
\;=\;
C_F\left(\frac{\mu^2 e^{-C}}{Q^2}\right)^s \nonumber\\
&&
\Bigg\{
  \frac{1}{s}
  \left[ \frac{5}{1+s} - \frac{\Gamma(1+s+n)}{\Gamma(1+s) \Gamma(1+n)}
        +\sum_{k=1}^n
         \frac{\Gamma(s+k)}{\Gamma(1+s) \Gamma(1+k)}
         \frac{4k}{1+s+k}
  \right]
\nonumber\\
&&+
  \frac{1}{1-s}
  \left[ \frac{4}{1+s} - \frac{2\Gamma(1+s+n)}{\Gamma(1+s) \Gamma(1+n)}
       + \sum_{k=1}^n
         \frac{\Gamma(s+k)}{\Gamma(1+s) \Gamma(1+k)}
         \frac{4(1+k)}{1+s+k}
  \right]
\nonumber\\
&&+
  \frac{1}{2-s}
  \left[ \frac{1}{1+s} + \frac{\Gamma(1+s+n)}{\Gamma(1+s) \Gamma(1+n)}
        +\sum_{k=1}^n
         \frac{\Gamma(s+k)}{\Gamma(1+s) \Gamma(1+k)}
         \frac{2}{1+s+k}
  \right]
\Bigg\}
\eeqar
where $s=\beta_0 u$ replaces the Borel parameter $u$.
We find IR-renormalon poles at $s=0,1,2$, as usual for DIS.
This is not a general statement
as in the case of $e^+e^-$-fragmentation one obtains
an infinite sum of
poles with even powers of $1/Q$ \cite{Das96}.
The Wilson coefficient has still to be renormalized
and the coefficient of the $1/s$-pole
should give the anomalous 
dimension as will be discussed in the next section.

In the Borel-plane the ambiguity $C_n^{(k)}$ 
of the perturbative series in
Eq.\,(\ref{truncate}) can be rediscovered as ambiguity of the inverse
Borel transformation due to the IR-renormalon poles
or in other words as the imaginary part of the Laplace integral:
\beq{inverse-borel}
\sum_k
C_n^{(k)} \left(\frac{\Lambda_C^2 e^{-C}}{Q^2}\right)^k
\;=\;
\frac{1}{\pi \beta_0}
{\rm Im} \int_0^\infty ds\;e^{-s/(\beta_0 a_s)}
{\cal B}_{\frac{1}{a_s}}
\left[C_n\left(\frac{Q^2}{\mu^2},a_s\right)\right](s)
\quad.
\eeq
We get
\beqar{ambiguity}
C_n^{(1)}
&=&
\pm \frac{2C_F}{\beta_0}
\left( 3+n - \frac{2}{1+n} - \frac{2}{2+n} - 2\sum_{k=1}^n\frac{1}{k}
\right)
\nonumber\\
C_n^{(2)}
&=&
\pm \frac{C_F}{\beta_0}
\left(  3 + \frac{1}{2}(1+n)(4+n)
      - \frac{2}{1+n} - \frac{2}{2+n} - \frac{2}{3+n}
      - 2\sum_{k=1}^n\frac{1}{k}
\right)
\quad.
\eeqar
The signs of these twist-2 uncertainty terms remain undetermined,
because it is not clear in which way the pole should be
circumvented in the Laplace integral.

The ambiguity of the Laplace integral is an intrinsic uncertainty of
the twist-2 transversity distribution. 
The ratio of the moments of this uncertainty $h_1^{IR}$ and the
moments of $h_1$ is expanded up to the
order $a_s/Q^2$
\beqar{momexpand}
M_n^{\rm IR}
&=&
\frac{C_n^{(1)} \frac{\Lambda^2 e^{-C}}{Q^2} 
      + {\cal O}\left( \frac{a_s}{Q^2}, \frac{1}{Q^4} \right)}
     {\sum_{k=0}^{m_0} B_n^{(k)} a_s^k
      + C_n^{(1)} \frac{\Lambda^2 e^{-C}}{Q^2}
      + {\cal O}\left( \frac{a_s}{Q^2}, \frac{1}{Q^4} \right)}
\, M_n
\nonumber\\
&\approx&
\left[
\frac{C_n^{(1)}}{B_n^{(0)}} \frac{\Lambda^2 e^{-C}}{Q^2}
+ {\cal O}\left( \frac{a_s}{Q^2}, \frac{1}{Q^4} \right)
\right]
M_n
\quad.
\eeqar
Here the truncated perturbative series Eq.\,(\ref{truncate}) was used
for the moments of $h_1$. The lowest order twist-2
perturbative coefficient $B_n^{(0)}$ is determined by the tree-graph
and is 1.
As an illustration we will insert for $M_n$ a
theoretical model prediction for the
transversity distribution.

>From Eq.\,(\ref{ambiguity}) and (\ref{momexpand}) we can calculate
the IR-renormalon uncertainty $U_n$ for each moment, where
\beq{momtw4}
\int_0^1 dx\,x^n 
\left(h_1(x)-(-)^n\overline{h}_1(x)\right)
\;=\; A_n^{(0)}\left(1\pm U_n\right)
+ {\cal O}(a_s)
\quad,
\eeq
where $A_n^{(0)}$ denotes the leading contribution to the
moments.
At $Q^2=4\,{\rm GeV}^2$ and with
$\Lambda_{\overline{MS}}=200\,{\rm MeV}$, $N_f=4$, $C_F=\frac{4}{3}$ and
$\beta_0=11-\frac{2}{3}N_f$ we find:
\beqar{mom04tw4}
&&
\int_0^1 dx \,
\left(h_1(x)-\overline{h}_1(x)\right)
\;=\;  A_0^{(0)}
+ {\cal O}(a_s)
\quad,
\nonumber\\
&&
\int_0^1 dx \,x
\left(h_1(x)+\overline{h}_1(x)\right)
\;=\;  A_1^{(0)} (1\pm 0.0057)
+ {\cal O}(a_s)
\quad,
\nonumber\\
&&
\int_0^1 dx \,x^2
\left(h_1(x)-\overline{h}_1(x)\right)
\;=\;  A_2^{(0)} (1\pm 0.014)
+ {\cal O}(a_s)
\quad,
\nonumber\\
&&
\int_0^1 dx \,x^3
\left(h_1(x)+\overline{h}_1(x)\right)
\;=\;  A_3^{(0)} (1\pm 0.024)
+ {\cal O}(a_s)
\quad,
\nonumber\\
&&
\int_0^1 dx \,x^4
\left(h_1(x)-\overline{h}_1(x)\right)
\;=\;  A_4^{(0)} (1\pm 0.036)
+ {\cal O}(a_s)
\quad.
\eeqar
We get no IR-renormalon uncertainty for the first moment.
The IR-renormalon uncertainty
becomes larger for higher moments, so that we expect larger 
ambiguities of the transversity distribution in the region of larger
Bjorken-$x$.
For higher moments the contribution of $\overline{h}_1$
can be considered as marginal, so that the above ambiguities should
remain approximately 
correct for $\int_0^1 dx x^n h_1(x)$ with $n>2$. 
This is of course not
the case for the first and the second moment ($n=1$).
A rough estimate gives rise to a sea-quark effect of the same 
order as the calculated uncertainty.

\begin{figure}[ht]
\vspace{1cm}
%\center{\hspace*{0mm} \epsfxsize=15cm \epsfysize10cm
%        \epsfbox{tw4_bag.eps}}
\centerline{\epsfig{figure=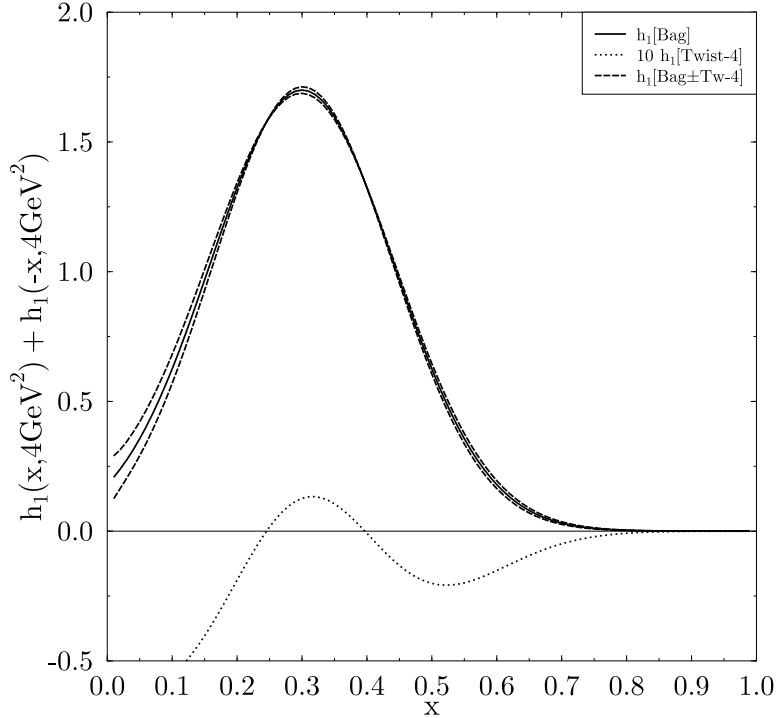,width=15cm}}
\caption[]{\sf
The bag model calculation \cite{Jaf91}
for $h_1-\overline{h}_1$ (full line) and the IR-renormalon ambiguity
evaluated at $Q^2=4\,{\rm GeV}^2$ multiplied by a factor of $10$
(dotted line). There are two changes in sign at $x\approx 0.25$ and
$x\approx 0.4$. The ambiguity was
added and subtracted from the
bag model calculation (dashed lines), 
($\La_{\overline{MS}}=200~{\rm MeV}$, 
and $N_f = 4$).}
\label{fig1}
\end{figure}
\begin{figure}[ht]
\vspace{1cm}
%\center{\hspace*{0mm} \epsfxsize=15cm \epsfysize10cm
%        \epsfbox{tw4_bagr.eps}}
\centerline{\epsfig{figure=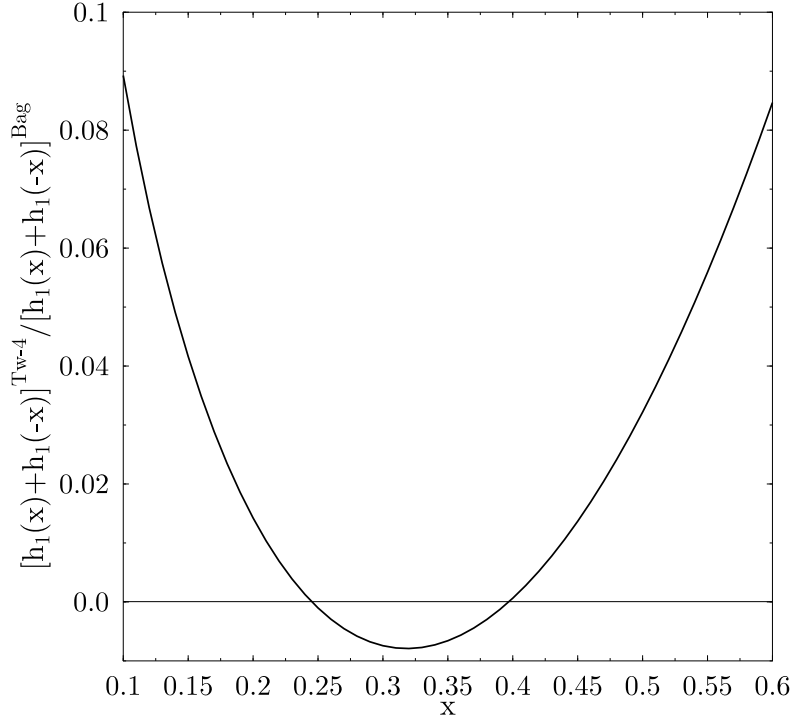,width=15cm}}
\caption[]{\sf
The relative magnitude of the IR-renormalon ambiguity with respect to
the bag model calculation \cite{Jaf91}
for $h_1-\overline{h}_1$
($\La_{\overline{MS}}=200~{\rm MeV}$, $Q^2=4\,{\rm GeV}^2$
and $N_f = 4$).}
\label{fig2}
\end{figure}
>From the moments in Eq.\,(\ref{momexpand}) the valence quark
transversity distribution
$h_1-\overline{h}_1$ may be reconstructed as a function of
Bjorken-$x$. The result is a convolution integral:
\beq{convolution}
h_1^{IR}(x,Q^2) - \overline{h}_1^{IR}(x,Q^2)
\;=\;
\frac{\Lambda^2 e^{-C}}{Q^2}
\int_x^1 \frac{dy}{y}\, \widetilde{C}^{(1)}(y)
\left\{ h_1\left(\frac{x}{y}\right)
     - \overline{h}_1\left(\frac{x}{y}\right)
\right\}
\eeq
where $\widetilde{C}^{(k)} (x)$ is defined by
\beq{Ctilde}
C_n^{(k)} \;=\; \int_0^1 dx\,x^n \widetilde{C}^{(k)}(x)
\eeq
and we find for the two first IR-renormalon ambiguities:
\beqar{C1tilde}
\widetilde{C}^{(1)}(x)
&=&
\pm \frac{2C_F}{\beta_0}
\left\{\frac{2}{(1-x)_+} + 3 \delta(x-1) - \delta'(x-1) - 2x -2
\right\}
\nonumber\\
\widetilde{C}^{(2)}(x)
&=&
\pm \frac{2C_F}{\beta_0}
\left\{  \frac{1}{(1-x)_+} 
       + \frac{5}{2}\delta(x-1) - \delta'(x-1) 
       + \frac{x}{4}\delta''(x-1)
       - x^2 - x - 1
\right\}
\;.
\nonumber\\
\eeqar

The IR-renormalon ambiguities
shown in Fig.\,\ref{fig1} are calculated using
the bag model calculation of \cite{Jaf91}
for $h_1(x)$.
%which is a relativistic model and for this reason
%should be sensitive to differences to the longitudinal twist-2 
%structure function $g_1$.
The large $N_f$-limit is most reliable in the
region of medium large Bjorken-$x$.
For small $x$ the neglection of multiple
gluon exchange is no longer justified. On
the other hand the influence of the hadronic spectrum makes a pure
perturbative calculation insufficient for large $x$.
In the region of best accuracy ($0.2<x<0.45$) the uncertainty does not
become bigger than 1\% (see Fig.\,\ref{fig2}).
One can expect a sizeable IR-renormalon uncertainty of up to 10\%
for $0.5<x<0.6$.

%==================================================================

\section{$\alpha_s$ corrections to the moments}

It is possible to extract the anomalous dimension and the
$\alpha_s$ corrections to the moments
of the nonsinglet
transversity distribution from Eq.\,(\ref{wilson}). This
expression has to be renormalized so that the Borel transformed
Wilson coefficient should be regularized simultaneously analytically
--- like it was done in the previous section --- and dimensionally.
We get for this generalized expression ($d=4-\ep$):
\beqar{sepswilson}
&&
\hspace*{-3mm} 
{\cal B}_{\frac{1}{a_s}}
\left[C_n\left(\frac{Q^2}{\mu^2},a_s\right)\right](s)
\;=\;
-C_F
\left(\frac{\mu^2 e^{-C}}{Q^2}\right)^s 
\left(\frac{4\pi\mu^2}{Q^2}\right)^{\frac{\ep}{2}}
\frac{ \Gamma\left(2-\frac{\ep}{2}\right)
       \Gamma\left(-\frac{\ep}{2}-s\right)}
     { \Gamma\left(3-\ep-s\right) \Gamma(1+s)}
\nonumber\\
&&
\hspace*{-3mm} 
\Bigg\{
\sum_{k=0}^n
  \frac{ \Gamma(1+s+k) \Gamma\left(s+k+\frac{\ep}{2}\right)}
       { \Gamma(1+k) \Gamma(2+s+k)}
  \left[   \left(s+\frac{\ep}{2}\right)
           \left( 6(1-s) + (4-\ep)(1-\ep)\right)
         + 4k (2-s-\ep) 
  \right]
\nonumber\\
&&
\hspace*{-3mm} 
- (2-\ep) \frac{ \Gamma\left(1+s+n+\frac{\ep}{2}\right)}
          { \Gamma(1+n)}
+ \ep s (2-s-\ep)
  \frac{ \Gamma\left(1+s+n+\frac{\ep}{2}\right)
         \Gamma(1+s+n)}
       { \Gamma(3+s+n) \Gamma(1+n)}
\Bigg\}
\quad.
\eeqar
It is important to take care of a possible violation of the Ward
identities due to the analytic continuation of $\gamma_5$ to
$d$-dimensions \cite{Lar91}.

The coefficients of the perturbative series can be reconstructed from
the Borel transformed Wilson coefficient by an appropriate number of
derivations with respect to $s$ at the origin $s=0$ of the Borel
plane:
\beq{ddsCn}
B_n^{(k+1)}
\;=\;
\left.
\beta_0^k \left(\frac{d}{ds}\right)^k
{\cal B}_\frac{1}{a_s} \left(C_n\right)(s) \right|_{s=0}
\quad.
\eeq
This may be verified starting with Eq.\,(\ref{truncate}) and using
\beq{borelas}
{\cal B}_\frac{1}{a_s}\left(a_s^{k+1}\right)(t)
\;=\;
\frac{t^k}{k{\rm !}}
\quad.
\eeq
So the first order $\alpha_s$ correction is determined by setting
$s=0$ in Eq.\,(\ref{sepswilson}). 
We find the quark-quark anomalous dimension of the nonsinglet
operator as the coefficient of the divergence
at $d=4$:
\beq{anomalous}
\gamma_{\perp,n}
\;=\;
\frac{C_F}{4\pi^2}
\left( \frac{1}{1+n} + \sum_{k=1}^n \frac{1}{k} \right)
\quad.
\eeq
This anomalous dimension
is identical with the coefficient of the
$1/s$-term in Eq.\,(\ref{wilson}) in the pure analytical
regularization, as it should be.
In momentum space we get the corresponding splitting function
\beq{Ph1}
P_\perp^{qq}(x)
\;=\;
\frac{C_F}{4\pi^2}
\left(1 - \frac{1}{(1-x)_+}\right)
\eeq
which is to be compared with the splitting function for
$h_1$ as found in the literature \cite{Art90}:
\beq{Pg1}
P_\perp^{qq}(x)
\;=\;
\frac{C_F}{4\pi^2}
\left(1 - \frac{1}{(1-x)_+}
- \frac{3}{4}\delta(x-1)
\right)
\quad.
\eeq
Both expressions differ by $\frac{3}{4} \delta(x-1)$.
This term is not really calculated in the conventional approach but
rather fixed by the requirement that the transversity flip
probability for quarks emitting a zero momentum gluon has
to vanish.\footnote{We thank X.\,Artru for helpful comments.}
In addition the $\delta$-function part should be the same as
for the polarized structure function $g_1$.
Following this standard argumentation the corresponding
$\delta(x-1)$-term has to
be added in Eq.\,(\ref{Ph1}). However, such additional arguments
should not be necessary in a direct calculation of the anomalous
dimension. We got the right $\delta$-function in the case of
$g_1$ without any additional arguments
in an analogous calculation \cite{Mey96}.
It is clear from a comparison of these two
calculations that the behaviour 
of the anomalous dimensions of $g_1$ and $h_1$
at $x\to 1$ may only be
indentical, if the vertex-correction with scalar- or vector-vertex
contributes in the same way for $x\to 1$. This is obviously not the
case, so that the difference of Eq.\,(\ref{Ph1}) and (\ref{Pg1})
at $x=1$ appears necessarily.
Considering the general arguments
leading to Eq.\,(\ref{Pg1}) as correct,
we conclude that the equivalence of the
twist-2 part of the used operator in Eq.\,(\ref{Tmu})
to $h_1$ leading to Eq.\,(\ref{Ph1}) is problematic
at $x=1$.
One should emphasize, however, 
that the renormalon estimation in the last
section is not affected of this question.

Unlike in the case of $g_1$
\cite{Mey96} the anomalous dimension does not vanish for any $n$,
so that at one loop level
a renormalization is already necessary for all moments.
We choose the $\overline{\rm MS}$-renormalization scheme. After
renormalization of the Ward-identity with
\beq{ward}
Z_5
\;=\;
1- C_F\, \frac{\alpha_s}{\pi} + {\cal O}(\alpha_s^2)
\eeq
we get for the first moments
of the structure function defined by Eq.\,(\ref{Tmu}):
\beqar{asmoment}
&&
\int_0^1 dx\,\left(h_1(x) - \overline{h}_1(x)\right)
\;=\;
1 + {\cal O}\left(\alpha_s^2\right)
\nonumber\\
&&
\int_0^1 dx\,x \left(h_1(x) + \overline{h}_1(x)\right)
\;=\;
1 + \frac{7}{8} C_F \frac{\alpha_s}{\pi} 
+ {\cal O}\left(\alpha_s^2\right)
\nonumber\\
&&
\int_0^1 dx\, x^2 \left(h_1(x) - \overline{h}_1(x)\right)
\;=\;
1 + \frac{13}{8} C_F \frac{\alpha_s}{\pi} 
  + {\cal O}\left(\alpha_s^2\right)
\nonumber\\
&&
\int_0^1 dx\, x^3 \left(h_1(x) + \overline{h}_1(x)\right)
\;=\;
1 + \frac{109}{48} C_F \frac{\alpha_s}{\pi} 
  + {\cal O}\left(\alpha_s^2\right)
\nonumber\\
&&
\int_0^1 dx\, x^4 \left(h_1(x) - \overline{h}_1(x)\right)
\;=\;
1 + \frac{227}{80} C_F \frac{\alpha_s}{\pi} 
  + {\cal O}\left(\alpha_s^2\right)
\quad.
\eeqar
As the used NNA-approximation is exact on the one-loop-level, these
first order perturbative corrections are exact results.

\section{Conclusions}

We calculated the IR-renormalon ambiguity for the valence quark
transversity distribution. The ambiguity
is smaller than 1\% in the region of best
accuracy of the NNA-approximation. For $x\approx 0.6$ this
systematic error becomes important reaching about 10\%.
Thus the renormalon uncertainty should be taken into accout when
interpreting measurements of $h_1$ in this region of $x$.
In addition we presented a calculation
of the anomalous dimension using the operator in Eq.\,(\ref{Tmu}) and
of the corresponding one loop perturbative corrections to the first
moments.
To our knowledge we
calculated the perturbative corrections for the first time.

\vspace*{5mm}
{\bf Acknowledgements.}
We thank L.\,Szymanowski and S.\,Hofmann for helpful discussions.
This work has been supported by  BMBF.
A.S.~thanks also the
MPI f\"ur Kernphysik in Heidelberg for support.

\vfill
\eject

\end{document}